\documentclass[twocolumn,english,prl]{revtex4}
\pdfoutput=1
\usepackage[T1]{fontenc}
\usepackage[latin9]{inputenc}
\usepackage{babel}
\usepackage{amsmath}
\usepackage{graphicx}
\usepackage{amssymb}
\usepackage{esint}
\usepackage[unicode=true, pdfusetitle,
 bookmarks=true,bookmarksnumbered=false,bookmarksopen=false,
 breaklinks=false,pdfborder={0 0 1},backref=false,colorlinks=false]{hyperref}

\makeatletter

\@ifundefined{textcolor}{}
{
 \definecolor{BLACK}{gray}{0}
 \definecolor{WHITE}{gray}{1}
 \definecolor{RED}{rgb}{1,0,0}
 \definecolor{GREEN}{rgb}{0,1,0}
 \definecolor{BLUE}{rgb}{0,0,1}
 \definecolor{CYAN}{cmyk}{1,0,0,0}
 \definecolor{MAGENTA}{cmyk}{0,1,0,0}
 \definecolor{YELLOW}{cmyk}{0,0,1,0}
 }
\makeatother
\begin{document}

\preprint{This line only printed with preprint option}
\title{Optical injection and terahertz detection of the macroscopic Berry
curvature}
\author{Kuljit S. Virk}
\email{kv2212@columbia.edu}
\affiliation{Department of Chemistry, Columbia University, 3000 Broadway, New York, USA,
10027}
\author{J.E. Sipe}
\email{sipe@physics.utoronto.ca}
\affiliation{Department of Physics and Institute for Optical Sciences, University of
Toronto, 60 St. George Street, Toronto, Ontario, Canada, M5S 1A7}

\begin{abstract}
We propose an experimental scheme to probe the Berry curvature of solids.
Our method is sensitive to arbitrary regions of the Brillouin zone, and
employs only basic optical and terahertz techniques to yield a background
free signal. Using semiconductor quantum wells as a prototypical system, we
discuss how to inject Berry curvature macroscopically, and probe it in a way
that provides information about the underlying microscopic Berry curvature.
\end{abstract}

\maketitle

Berry's phase permeates many fields of physics. In quantum mechanics, the
net phase acquired by a wavefunction during a cyclic change in the
parameters of a Hamiltonian is a gauge invariant quantity that has
measurable effects. This quantity can be directly expressed
in terms of a physical property, called the \emph{Berry curvature }(BC),
which relates to the Berry phase in much the same way as the gauge-invariant
magnetic field relates to a gauge-dependent vector potential \cite%
{Wilczek}. \ In materials science, it is rapidly becoming clear
that the role of the BC is as fundamental as that of an energy band \textsc{%
\cite{Martin}.} The representation of position operator in terms of the
Bloch states of a crystal is intimately related to the BC, and it thus
appeared in the works of Adams and Blount \cite{Blount1962305} on
the topic much earlier than the rigorous formulation by Berry \cite%
{BerryOriginalPaper}. In 1990s, King-Smith and Vanderbilt introduced a
theory of electric polarization of solids as a bulk quantity \cite%
{Vanderbilt1993}, the present version of which is grounded in Berry's phase 
\cite{Martin,Resta1994}. In parallel to this, the BC has appeared as a
central physical quantity in the work on anomalous Hall effect (AHE) in
ferromagnetic materials \cite{onoda:165103,Jungwirth2002}, and the intrinsic
mechanism of spin Hall effect (SHE) \cite{Murakami:2006p1740,Engel2008}.
Haldane expressed the non-quantized part of the intrinsic Hall
conductivity in terms of integral of the \emph{Berry connection} (vector potential of BC) on the Fermi surface \cite{Haldane:2004p1736}, thus recapturing the essence of Landau's Fermi liquid theory. Recent works also show the role of BC in the photogalvanic effect \cite{Deyo,Hosur}. 

Central to these effects is \emph{anomalous velocity}, which refers to
motion of charges perpendicular to their usual group velocity \cite%
{Blount1962305}. As an average of the BC it appears as its main experimental
manifestation. Though the above works
elucidate the role of the BC in \emph{explaining} various phenomena, they
are severely restricted as methods to \emph{probe} it as a \emph{basic
property} of solids. This is because the d.c. response reduces to an integral of Berry curvature over the full BZ, and for partially filled band, to the volume occupied by the carrier distribution function \cite{Haldane:2004p1736}. The measured response is thus restricted either to the full BZ, or slightly displaced equilibrium occupation function of a partially filled band. 

\begin{figure}[tbp]
\begin{centering}
\includegraphics[width=2.5in]{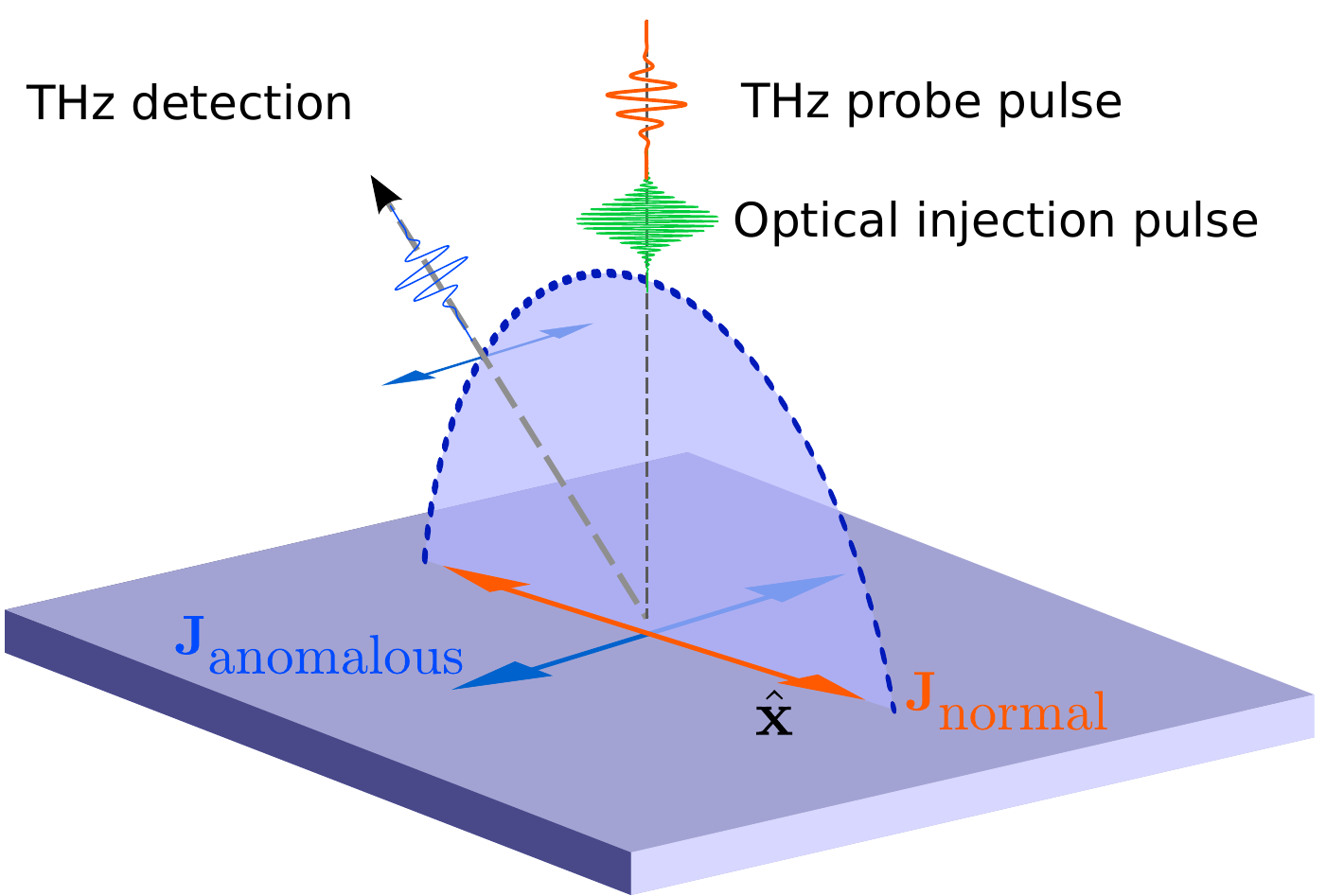}
\par\end{centering}
\caption{Illustration of the proposed scheme for injection and detection of
macroscopic Berry curvature}
\label{fig:Illustration-of-the}
\end{figure}

This letter presents a method to study this important and complementary
quantity of solids, which is much less restricted in its sampling of the Brillouin
zone. It was anticipated by the 
work of Moore and Orenstein\cite{Moore}, who showed how confinement in quantum wells induces a Berry's phase that leads to a helicity dependent photocurrent. While we focus on the 
hole bands of a GaAs quantum well (see Fig. \ref{fig:Illustration-of-the}), chosen because of their large 
BC \cite{Murakami:2006p1740} with a rich structure, our method is very general and in principle can be 
applied to a wide range of materials. An optical
excitation by \emph{circularly polarized} light induces a non-zero transient
macroscopic Berry curvature (TMBC) in such materials, due to the creation of
a state that breaks time-reversal and space inversion symmetry. The
injection is followed by a \emph{linearly polarized} terahertz wave, which
drives the optically injected carriers. In the presence of TMBC, the motion
of charges has an anomalous component, which is perpendicular to the
polarization of the terahertz wave. The detection of the radiated THz field
in the cross-polarized direction is thus \emph{entirely} a signal due to the
TMBC. Such a TMBC can be injected even in materials of Chern class zero,
where an equilibrium MBC is not allowed \citep{Hasan2010}. This is in contrast to spin hall insulators in which a net MBC exists in the ground state.

Our letter is organized as follows. We begin with a discussion of the
microscopic Berry curvature. We then discuss the optical injection of TMBC,
its detection by linearly polarized THz excitation, and its intrinsic
lifetime due to electron-hole scattering. Finally, we present the results of
our numerical calculations of the microscopic and optically injected TMBC,
followed by the results of the anomalous velocity induced by the THz
excitation.

\section{Microscopic Berry Curvature}

\textbf{Bulk crystals:} The band theory of solids describes single electron
wavefunctions by Bloch functions, $\psi _{n}(\boldsymbol{k};\boldsymbol{r})$%
, for energy bands, $\varepsilon _{n}(\boldsymbol{k})$. Here $n$ labels a
single energy band with a continuous slope, starting from the energy levels
at the center of the Brillouin zone \cite{lax}. Between two degenerate
bands, or within a single band, the vector 
\begin{equation}
\boldsymbol{\zeta }_{nm}(\boldsymbol{k})=i\int d\boldsymbol{r}\,u_{n}^{\ast
}(\boldsymbol{k};\boldsymbol{r})\frac{\partial }{\partial \boldsymbol{k}}%
u_{m}(\boldsymbol{k};\boldsymbol{r})\label{eq:zeta def},
\end{equation}%
acts like a vector potential in the momentum space dynamics of electrons,
its dependence on the choice of the phases of the $\left\{ u_{m}(\boldsymbol{%
k};\boldsymbol{r})\right\} $ leading to the analogue of a gauge dependence.
Taking the curl of $\boldsymbol{\zeta }_{nm}(\boldsymbol{k})$ leads to the
(microscopic) BC, a "gauge invariant" quantity at each $\mathbf{k}$,%
\begin{equation}
\Omega _{nm}^{\alpha }(\boldsymbol{k})=(\nabla \times \zeta _{nm}(%
\boldsymbol{k}))^{\alpha }-i\epsilon _{\alpha \beta \gamma }\sum_{\varepsilon
_{p}=\varepsilon _{n}=\varepsilon _{m}}\left[ \zeta _{np}^{\beta }(%
\boldsymbol{k}),\,\,\zeta _{pm}^{\gamma }(\boldsymbol{k})\right] ,
\label{eq:Omega}
\end{equation}%
In the presence of time-reversal and spatial inversion symmetries, $%
\boldsymbol{\Omega }(\boldsymbol{k})$ takes the form of a traceless matrix
within each degenerate subspace \cite{Blount1962305}; the macroscopic Berry
curvature (MBC), 
\begin{equation*}
\left\langle \boldsymbol{\Omega }\right\rangle \equiv \sum_{nm\boldsymbol{k}}%
\boldsymbol{\Omega }_{nm}(\boldsymbol{k})\rho _{mn}(\boldsymbol{k}),
\end{equation*}%
where $\rho _{nm}(\mathbf{k})$ is the single particle density matrix, then
vanishes in equilibrium; even in a material such as GaAs, where there is no
inversion symmetry, the equilibrium MBC vanishes because the band structure
is of Chern class zero.   Away from equilibrium, such as in the d.c.
response of a $p$-doped semiconductor, the MBC can play a role \cite%
{Murakami:2006p1740,Jungwirth2002,Engel2008}, but optically excited distributions allow much greater access to its local probing in \textbf{k}-space.

\textbf{Quantum wells:} We focus on the valence states of a quantum well,
described by the Luttinger model of a square quantum well grown along the $%
\left[ 001\right] \equiv \hat{\mathbf{z}}$ direction. In this model  \cite%
{Andreani1987} there are degenerate ($\pm $) wavefunctions described by
4-component spinors, labeled $\boldsymbol{f}_{n}^{\pm }(\mathbf{k};z)$ for
each 2-dimensional subspace $n$, and the microscopic BC is equal to $\mathbf{%
\hat{z}}\Omega _{nn}^{z}(\boldsymbol{k})\sigma ^{3}$, where%
\begin{equation}
\Omega _{nn}^{z}(\boldsymbol{k})=\epsilon _{z\mu \nu }\int dz\partial _{\mu }\boldsymbol{f}_{n}^{-\dagger }(\boldsymbol{k};z)\cdot
\partial _{\nu }\boldsymbol{f}_{n}^{-}(\boldsymbol{k};z),  \label{eq:Omega n}
\end{equation}%
and $\sigma ^{3}$ is the third Pauli matrix. Figure \ref%
{fig:Energy-bands-(solid)} shows the top two valence energy bands and the
corresponding Berry curvature for a 15 nm thick, quantum well. The middle
panel shows $\Omega _{nn}^{z}(k)$ defined in \eqref{eq:Omega n}; large $%
\Omega _{nn}^{z}(k)$ results from a large mixing of two or more states%
\footnote{%
To verify, use $\boldsymbol{k}\cdot \boldsymbol{p}$ expansion and \eqref{eq:zeta def}
and \eqref{eq:Omega}}, here arising because of contributions from the light
hole to the two bands.

\begin{figure*}[tbp]
\begin{minipage}[t]{0.2\textwidth}%
\begin{center}
\includegraphics[scale=0.65]{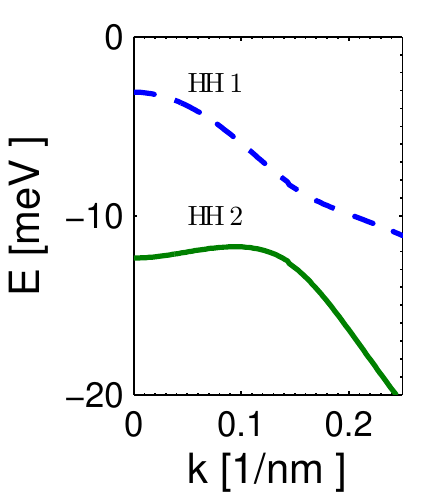}\\(a)
\par\end{center}%
\end{minipage}\hfill{}%
\begin{minipage}[t]{0.2\textwidth}%
\begin{center}
\includegraphics[scale=0.65]{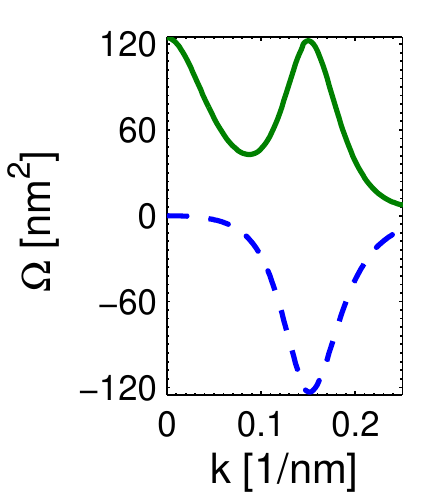}\\(b)
\par\end{center}%
\end{minipage}\hfill{}%
\begin{minipage}[t]{0.2\textwidth}%
\begin{center}
\includegraphics[scale=0.65]{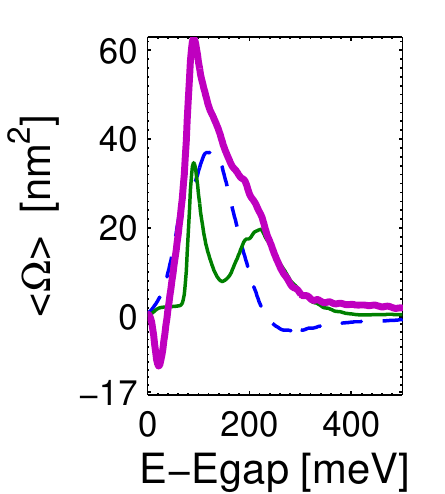}\\(c)
\par\end{center}%
\end{minipage}\hfill{}%
\begin{minipage}[t]{0.2\textwidth}%
\includegraphics[scale=0.75]{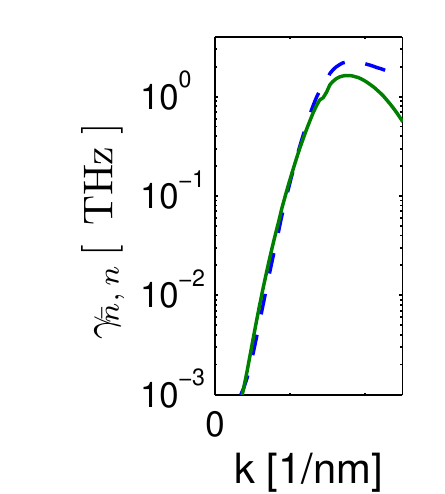}\\(d)%
\end{minipage}
\caption{(a) Highest valence (heavy) hole bands for a 15 nm thick {[}%
001{]} GaAs quantum well; (b) Berry curvature $\Omega^{z}(k)$ (lines as in (a)); (c) $%
\left\langle \Omega^{z}\right\rangle $ (thick solid line) and $\left\langle
\Omega_{n}^{z}\right\rangle $ for odd parity states (lines as in (a)); (d) Coulomb scattering rate
between the $+$ and $-$ states for optical excitation
at 90 meV above gap energy. Lines as in (a) }
\label{fig:Energy-bands-(solid)}
\end{figure*}

\section{Macroscopic Berry curvature}

No macroscopic effect results from $\Omega _{nn}^{z}(\boldsymbol{k})$ unless
an imbalance is created between the hole populations of the $\boldsymbol{f}%
^{\pm }$ states of these subspaces, leading to a nonzero $\left\langle
\Omega ^{z}\right\rangle $. In optical excitation across the bandgap with
left- circularly polarized light, the matrix elements connect $\boldsymbol{f}%
^{-}$ to the spin up conduction subbands with a probability $1/3$ of the
corresponding transition from $\boldsymbol{f}_{n}^{+}$ to the same
conduction subband. This yields a 3:1 population imbalance between the two
otherwise degenerate states, leading to a non-vanishing $\left\langle \Omega
^{z}\right\rangle $. The possibility of creating this imbalance in the non-equilibrium state is a well-established experimental fact \citep{AliPaper}. A correction arises from the electron-hole interaction such that the final state is an exciton. This correction is  small for ionization states (the only states relevant above the optical gap), and decreases the higher up we move in the band.

To access $\left\langle\Omega ^{z}\right\rangle $ we imagine a \emph{linearly polarized} THz electric field, $E(t)%
\hat{\mathbf{x}}$, driving the optically injected hole population. Terahertz
excitation does not couple $\boldsymbol{f}_{n}^{+}$ and $\boldsymbol{f}%
_{n}^{-}$ states since there is no momentum matrix element between them, and
therefore it does not disturb the injected $\left\langle \Omega
^{z}\right\rangle $. The Schrödinger equation for the hole density matrix $%
\rho (\mathbf{k},t)$ coupled to the external THz field polarized along $\hat{%
\mathbf{x}}$ (or {[}100{]}) direction reads,%
\begin{eqnarray}
&&i\left( \hbar \frac{\partial }{\partial t}+eE^{x}(t)\frac{\partial }{%
\partial k_{x}}\right) \rho (\boldsymbol{k},t)  \label{eq:EOM} \\
&=&\left[ H_{0}(\boldsymbol{k})+eE^{x}(t)\zeta ^{x}(\boldsymbol{k}),\rho (%
\boldsymbol{k},t)\right] +\left. \dot{\rho}\right\vert _{scatt}.  \notag
\end{eqnarray}%
Here $H_{0}(\boldsymbol{k})$ is the Luttinger Hamiltonian, and the last term
is the contribution of Coulomb scattering. A gauge transformation exists
such that $\zeta ^{x}(\boldsymbol{k})=0$ locally in the region of interest,
and within this gauge we can easily compute the solution of \eqref{eq:EOM}
with $\rho (\boldsymbol{k},0)$ set equal to the hole populations created by
optical excitation. The expectation value of the velocity operator, $\mathbf{%
v}$, is then calculated at each time point to study the macroscopic motion
of charges, 
\begin{equation}
\left\langle \boldsymbol{v}(t)\right\rangle =\sum_{\boldsymbol{k}}\rho _{nm}(%
\boldsymbol{k},t)\mathbf{v}_{mn}(\boldsymbol{k}).  \label{eq:velocity}
\end{equation}%
Sum rules that relate $\boldsymbol{\Omega }$ to $\boldsymbol{\zeta }$ \cite%
{lax} can be used to show that in the linear regime an \emph{%
anomalous} contribution to the average velocity may exist, and is
proportional to the macroscopic Berry curvature, 
\begin{equation}
\left\langle \boldsymbol{v}(t)\right\rangle _{anomalous}=-\hat{\mathbf{y}}%
\frac{e}{\hbar }\left\langle \Omega ^{z}\right\rangle E^{x}(t)\text{,}
\label{anomalous}
\end{equation}%
until $\left\langle \Omega ^{z}\right\rangle $ decays due to scattering
between the $\pm $ state populations, primarily through the dynamic
polarizability of the electron gas, described by the term $\left. \dot{\rho}%
\right\vert _{scatt}$. The electrons generated by the optical excitation would always diminish the Berry phase effects, and therefore this is an intrinsic lifetime because it survives in
the limit of no impurities. The self energy effects may be decomposed into
two contributions: (A) decoherence within each set (either $+$ or $-$) of
states, and (B) population transfer between the sets of states. Type A limit
the sensitivity of the dynamics within each set to the microscopic BC, while
type B reduce the net effect of MBC; B effects degrade the signal more
effectively than A effects by restoring the time-reversal symmetry broken by
the initial optical excitation.

We model these effects by setting $\left. \dot{\rho}_{nm}(\boldsymbol{k}%
,t)\right\vert _{scatt}=-\eta _{nm}\rho _{nm}(t)$ for off-diagonal terms
(type A; $n$ and $m$ in the same set), and $\left. \dot{\rho}_{nn}(%
\boldsymbol{k},t)\right\vert _{scatt}=-\gamma _{n\bar{n}}(\boldsymbol{k})%
\left[ \rho _{nn}(\boldsymbol{k},t)-\rho _{\bar{n}\bar{n}}(\boldsymbol{k},t)%
\right] $ for population relaxation terms (type B; here $n$($\bar{n}$)
refers to the \textbf{$\boldsymbol{f}_{n}^{+}$}($\boldsymbol{f}_{n}^{-}$)
state). To estimate the rates $\gamma _{nm}(\boldsymbol{k})$ and $\gamma _{n%
\bar{n}}(\boldsymbol{k})$, we start with the RPA self-energy, employ the
generalized Kadanoff Baym ansatz \cite{lipavsky1986gkb}, and follow by the
Markov approximation. The Auger scattering of holes by electrons is
negligible, due to the energy conservation condition at each vertex in the
self energy diagram.

\section{Results and discussion}

For excitation with a $100$ fs Gaussian optical pulse, the resulting $%
\left\langle \Omega ^{z}\right\rangle $ is shown as a function of energy
above the band gap in the third panel of Fig. \ref{fig:Energy-bands-(solid)}%
. The panel also shows the breakdown of $\left\langle \Omega
^{z}\right\rangle $ into contributions from the individual bands. The
initial rise in $\left\langle \Omega ^{z}\right\rangle $ is entirely due the
first band, in which the dominance of $\boldsymbol{f}^{+}$ component leads
to the sign opposite to its microscopic counterpart in the middle panel. The
large contribution from the second band starts to dominate above 85 meV.

The rightmost panel in figure \ref{fig:Energy-bands-(solid)} shows the
scattering rate $\gamma _{n,\bar{n}}(k)$ for each of the two bands at a
carrier density of $10^{11}\mbox{ cm}^{-2}$. \ The rate remains below 1 THz,
and is suppressed at small $k$, where the spinors have orthogonal dominant
contributions. We find that the $\eta _{nm}(k)$ range between 1 to 4 THz,
but their effect on the THz dynamics is only secondary as discussed in
relation to type A interactions. So we can expect that $\left\langle
\Omega ^{z}\right\rangle $ does not vanish during the subsequent THz probe. Thus we see that the effects of relevant scattering between electrons and holes are significantly suppressed by symmetries of the well. 

We now turn to the results of our numerical solution of \eqref{eq:EOM}, and
present the normal and anomalous hole velocities calculated via %
\eqref{eq:velocity}. The field is polarized along $\hat{\mathbf{x}}$ so that
the anomalous velocity is along $\hat{\mathbf{y}}$; see Figure 1. Our
calculations are performed with $E(t)=E_{0}e^{-(t-t_{0})^{2}/2\tau
_{p}^{2}}\cos (\omega _{0}t)$. We chose $E_{0}=0.1$ kV/cm, $\omega _{0}=1$
THz, $\tau _{p}=1$ ps, and $t_{0}$ a conveniently chosen point in time. By
comparing the velocities with and without the $\left. \dot{\rho}\right\vert
_{scatt}$ term in \eqref{eq:EOM}, we found that the the anomalous velocity
is reduced by $\simeq 36\%$ while the normal velocity undergoes little
change by scattering as expected.

\begin{figure}[tbp]
\begin{minipage}[t]{4.2cm}%
\includegraphics[width=4cm,height=3cm]{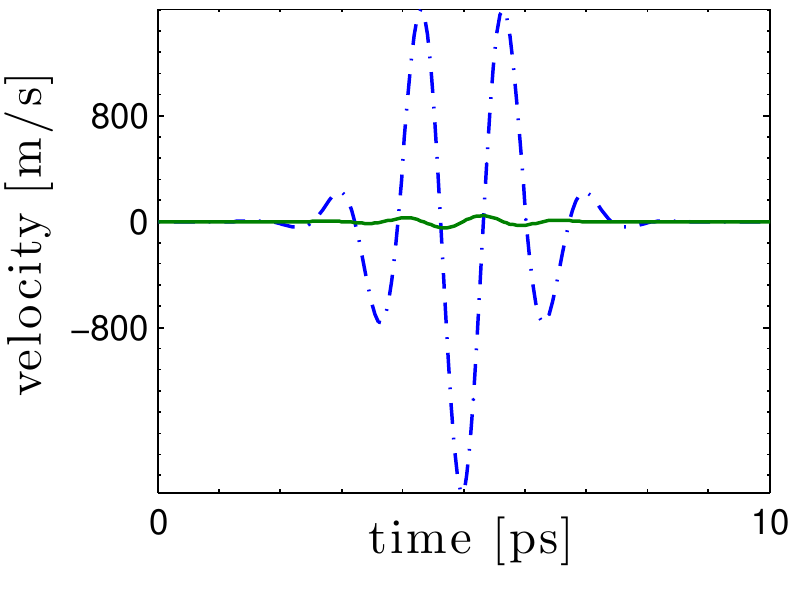}%
\end{minipage}\qquad{}%
\begin{minipage}[t]{2.6cm}%
\includegraphics[width=2.6cm,height=3cm]{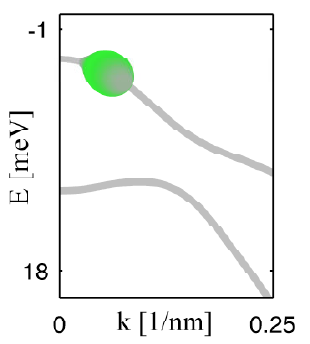}
\end{minipage}
\par
\medskip{}
\par
\begin{minipage}[t]{4.2cm}%
\includegraphics[width=4cm,height=3cm]{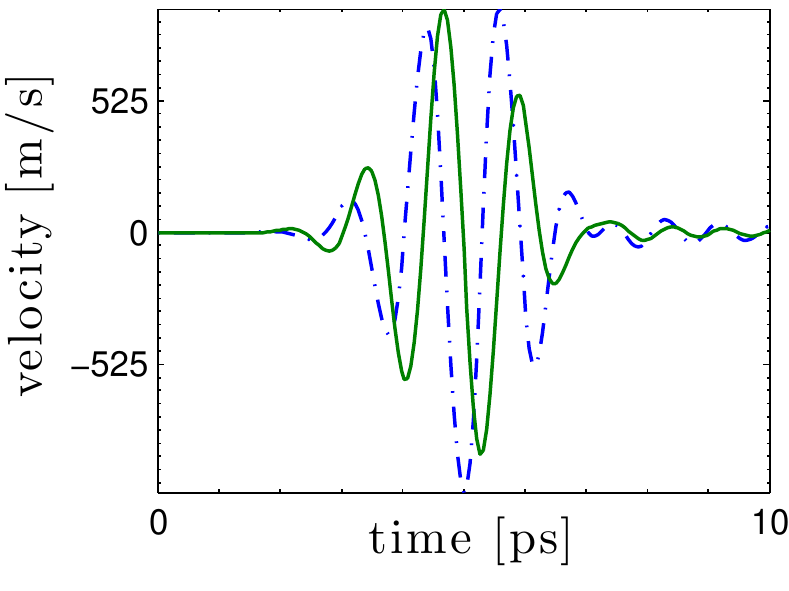}%
\end{minipage}\qquad{}%
\begin{minipage}[t]{2.6cm}%
\includegraphics[width=2.6cm,height=3cm]{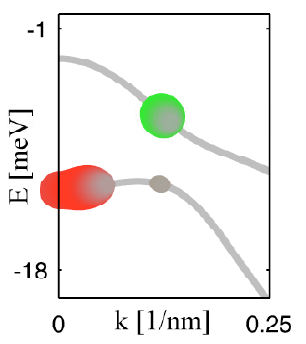}
\end{minipage}
\par
\medskip{} 
\begin{minipage}[t]{4.2cm}%
\includegraphics[width=4cm,height=3cm]{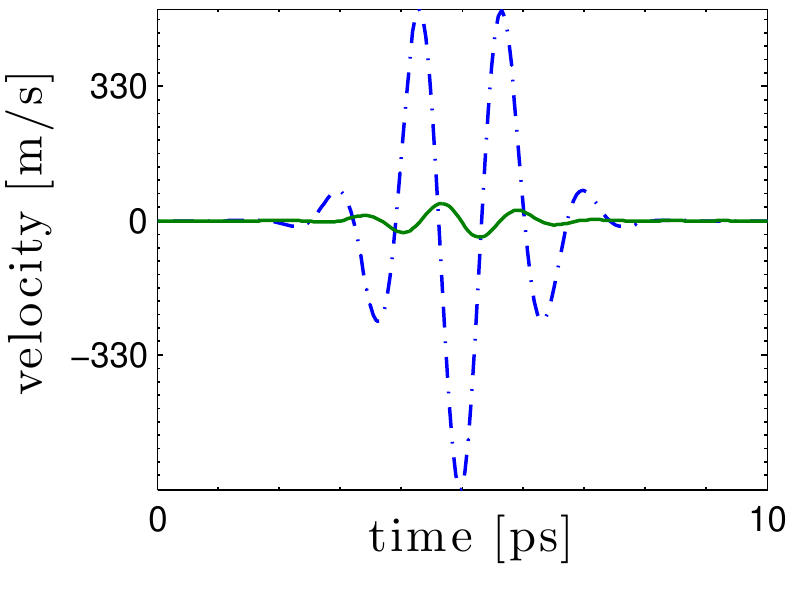}%
\end{minipage}\qquad{}%
\begin{minipage}[t]{2.6cm}%
\includegraphics[width=2.6cm,height=3cm]{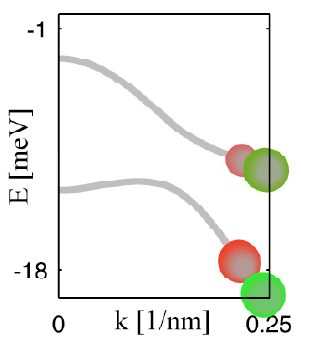}
\end{minipage}
\caption{THz induced velocities (left) and populations (right) at 20 meV, 90
meV, and 350 meV (top-bottom) for 15 nm wide QW. Anomalous (normal)
velocities are in solid (dashed-dotted) lines. }
\label{fig:Velocities-1}
\end{figure}

In Figure \ref{fig:Velocities-1} we demonstrate the relationship between the
magnitude of $\left\langle v_{y}(t)\right\rangle $, $\left\langle \Omega
^{z}\right\rangle $, and the hole populations for different photon energies
above the gap; large anomalous velocities clearly arise when the Berry
curvature of the populated states is large. These calculations lie in the
linear regime, where (\ref{anomalous}) is a good approximation; in the
non-linear regime, no clear cut relationship exists between the expectation
value of the velocity operator and the $\left\langle \Omega
^{z}\right\rangle $ as the concept of intraband motion itself breaks down.

As shown in Fig. \ref{fig:Illustration-of-the}, the anomalous velocity could
be detected by measuring the emitted THz radiation polarized perpendicular
to the incident field. While the much lighter electrons in the conduction
band would also emit THz radiation, and at much higher power than the holes,
the Berry curvature of conduction bands in these quantum wells vanishes. So
only\emph{\ }the\emph{\ }anomalous velocity of holes would contribute to the
THz emission perpendicular to the incident field. In addition, placing the
THz detector at an oblique angle in the plane of the incident field vector,
the dipole field of the parallel component can be highly suppressed while
leaving the perpendicular component unaffected. Separation of the desired
signal from background is thus already built into this method.

In the above analysis, we only took intrinsic scattering due to Coulomb
interaction among electrons into account. We emphasize that a large hole
mobility is crucial to the experimental success of this scheme. Spin-flip
scattering of holes is an approximate concept in this scenario, but it is a
useful characterization of the detrimental effects of impurity scatterers in
our scheme. It is suppressed exponentially close to the $\Gamma $ point, but
rises sharply to rates faster than 1 THz as a function of $k$ for areal
concentrations of $10^{10}\mbox{ cm}^{-2}$ \cite{Ferreira1991}. The effects
of phonons are subtle. On the one hand, they may provide a dominant
scattering channel. On the other hand, as observed previously, only those
scattering events that link the partner states are actually detrimental.
Detailed calculations of these effects, including lattice vibrations, is the
focus of our ongoing work.

In conclusion, we have shown that significant macroscopic Berry curvature
can be injected in GaAs quantum wells by circularly polarized light. The
lifetime of this macroscopic effect is at least a few picoseconds, and its
robustness results from the properties of hole wavefunctions under the
symmetry operations of the quantum well. We have presented a scheme to make
this macroscopic effect accessible experimentally via the anomalous
contribution to the THz emission from a quantum well driven by a linearly
polarized THz field. A successful implementation of this scheme would open
up a new venue in exploration of the Berry curvature as a fundamental
property of solids.

The authors acknowledge financial support from Natural Sciences and Research
Council of Canada. We thank Ali Najmaie and Sangam Chatterjee for insightful
discussions.

\bibliography{BerryCurvature}

\end{document}